\documentclass[twocolumn,showpacs,preprintnumbers,amsmath,amssymb,prl]{revtex4}

\usepackage{graphicx}
\usepackage{dcolumn}
\usepackage{bm}
\usepackage{tabularx}
\usepackage{amsmath}
\usepackage{textcomp}
\usepackage{upgreek}

\begin{document}

\title{Universal Heat Conduction in Ce$_{1-x}$Yb$_x$CoIn$_5$: Evidence for Robust Nodal $d$-wave Superconducting Gap}

\author{Y. Xu,$^1$ J. K. Dong,$^{1,2,*}$ L. I. Lum,$^3$ J. Zhang,$^1$ X. C. Hong,$^1$ L. P. He,$^1$ K. F. Wang,$^{4,\S}$ Y. C. Ma,$^{4,\P}$ C. Petrovic,$^4$ M. B. Maple,$^3$ L. Shu,$^{1,5,\dag}$ and S. Y. Li$^{1,2,5,\ddag}$ }

\affiliation{$^1$State Key Laboratory of Surface Physics and Department of Physics, Fudan University, Shanghai 200433, P. R. China\\
$^2$Laboratory of Advanced Materials, Fudan University, Shanghai 200433, P. R. China\\
$^3$Department of Physics, University of California, San Diego, La Jolla, California 92093, USA\\
$^4$Condensed Matter Physics and Materials Science Department, Brookhaven National Laboratory, Upton, New York 11973, USA\\
$^5$Collaborative Innovation Center of Advanced Microstructures, Fudan University, Shanghai 200433, P. R. China}

\date{\today}

\begin{abstract}
In heavy-fermion superconductor Ce$_{1-x}$Yb$_x$CoIn$_5$ system, Yb doping was reported to cause a possible change from nodal $d$-wave superconductivity to a fully gapped $d$-wave molecular superfluid of composite pairs near $x \approx$ 0.07 (nominal value $x_{nom}$ = 0.2). Here we present systematic thermal conductivity measurements on Ce$_{1-x}$Yb$_x$CoIn$_5$ ($x$ = 0.013, 0.084, and 0.163) single crystals. The observed finite residual linear term $\kappa_0/T$ is insensitive to Yb doping, verifying the universal heat conduction of nodal $d$-wave superconducting gap in Ce$_{1-x}$Yb$_x$CoIn$_5$. Similar universal heat conduction is also observed in CeCo(In$_{1-y}$Cd$_y$)$_5$ system. These results reveal robust nodal $d$-wave gap in CeCoIn$_5$ upon Yb or Cd doping.

\end{abstract}

\maketitle


CeCoIn$_5$ is an archetypal heavy-fermion superconductor with superconducting transition temperature $T_c$ $\simeq$ 2.3 K \cite{petrovic}. With many similarities to the high-$T_c$ cuprates, including quasi-two-dimensionality (quasi-2D), proximity to antiferromagnetism, and non-Fermi-liquid normal state, the unconventional superconductivity in CeCoIn$_5$ has long been thought to be due to spin fluctuation pairing \cite{Norman,Scalapino,Aynajian,PNAS2014}. The superconducting gap structure $\Delta(k)$ of CeCoIn$_5$ was widely studied by various experimental probes ever since it was discovered. These include specific heat \cite{An}, thermal conductivity \cite{Movshovich,Louis,Izawa,Flouquet}, as well as surface-sensitive techniques such as point contact spectroscopy and scanning tunneling microscopy \cite{Greene,Davic,Yazdani}. A $d$-wave superconducting gap with symmetry-protected nodes has been well established in CeCoIn$_5$.

The investigation of impurity effects can give a better understanding of the exotic normal and superconducting state of CeCoIn$_5$ \cite{Paglione,PNAS2011}. Recently, the anomalous phenomena observed in Ce$_{1-x}$Yb$_x$CoIn$_5$ system have attracted much attention \cite{Shu,Shimozawa,Jang,Polyakov,Dudy,PNAS2013,Kim,Coleman,Booth,Mizukami,White}. At first, the violation of Vegard's law was found in Ce$_{1-x}$Yb$_x$CoIn$_5$ single crystals, together with the robustness of $T_c$ and Kondo-coherence temperature $T_{coh}$ upon Yb doping \cite{Shu}. However, later measurements on the Ce$_{1-x}$Yb$_x$CoIn$_5$ thin films demonstrated a verification of Vegard's law, and strong suppression of $T_c$ and $T_{coh}$ with Yb doping \cite{Shimozawa}. To solve this discrepancy, Jang {\it et al.} carefully determined the actual Yb concentration $x_{act}$ in Ce$_{1-x}$Yb$_x$CoIn$_5$ single crystals, and found that $x_{act}$ is only about 1/3 of the nominal value $x_{nom}$ up to $x_{nom} \approx$ 0.5 \cite{Jang}. With $x_{act}$, the rate of $T_c$ suppression with Yb concentration for the single crystals is nearly the same as that observed in the thin films \cite{Jang}.

Nevertheless, the remarkable anomalies observed at $x_{nom}$ = 0.2 are still very puzzling, including Fermi surface topology change \cite{Polyakov}, Yb valence transition \cite{Dudy}, significant quasiparticle effective mass reduction as well as suppression of the quantum critical point \cite{Polyakov,PNAS2013}. Moreover, a recent London penetration depth study by Kim {\it et al.} suggested the nodal $d$-wave superconductivity becomes fully gapped beyond the critical Yb doping $x_{nom} = 0.2$ \cite{Kim}. To explain it, an exotic scenario was proposed, in which the nodal Fermi surface undergoes a Lifshitz transition upon Yb doping, forming a fully-gapped $d$-wave molecular superfluid of composite pairs \cite{Coleman}.

To examine such an exotic scenario, more experiments are highly desired to investigate the superconducting gap structure of Ce$_{1-x}$Yb$_x$CoIn$_5$ system. Low-temperature thermal conductivity measurement is an established bulk technique to probe the gap structure of a superconductor \cite{HShakeripour}. According to the magnitude of residual linear term $\kappa_0/T$ in zero field, one may judge whether there exist gap nodes or not. The field dependence of $\kappa_0/T$ can give further information on nodal gap, gap anisotropy, or multiple gaps.

In this Letter, we report a systematic heat transport study of Ce$_{1-x}$Yb$_x$CoIn$_5$ ($x_{nom}$ = 0.05, 0.2, and 0.4) single crystals. Finite $\kappa_0/T$ is observed in all three samples, which does not support a fully-gapped superconducting state at $x_{nom} \geq 0.2$. Furthermore, $\kappa_0/T$ manifests a nearly constant value upon doping, $i.e.$, the universal heat conduction, which is an important property of nodal $d$-wave superconducting gap. Similar universal heat conduction is also observed in CeCo(In$_{1-y}$Cd$_y$)$_5$ ($y_{nom}$ = 0.05, 0.075, and 0.1) system. These results demonstrate that the nodal $d$-wave gap in CeCoIn$_5$ is robust against Yb or Cd doping.

High-quality Ce$_{1-x}$Yb$_x$CoIn$_5$ ($x_{nom}$ = 0.05, 0.2, and 0.4) and CeCo(In$_{1-y}$Cd$_y$)$_5$ ($y_{nom}$ = 0.05, 0.075, and 0.1) single crystals were grown by a standard indium self-flux method \cite{Zapf,Pham}. Samples were etched in dilute hydrochloric acid to remove the In flux on the surfaces. The actual Yb concentration $x_{act}$ = 0.013, 0.084, and 0.163, and the actual Cd concentration $y_{act}$ = 0.004, 0.008, and 0.011 were determined by wavelength-dispersive spectroscopy (WDS), utilizing an electron probe microanalyzer (Shimadzu EPMA-1720H). Hereafter $x$ and $y$ represent $x_{act}$ and $y_{act}$, respectively. The samples were cut and polished into rectangular shape. Contacts were made with soldered indium for Ce$_{1-x}$Yb$_x$CoIn$_5$ and spot welding for CeCo(In$_{1-y}$Cd$_y$)$_5$, respectively, which were used for both in-plane resistivity and thermal conductivity measurements. The resistivity measurements were done in a $^3$He cryostat. The thermal conductivity was measured in a dilution refrigerator, using a standard four-wire steady-state method with two RuO$_2$ chip thermometers, calibrated $in$ $situ$ against a reference RuO$_2$ thermometer. Magnetic fields were applied along the $c$ axis and perpendicular to the heat current. To ensure a homogeneous field distribution in the samples, all fields were applied at a temperature above $T_c$.


\begin{figure}
\includegraphics[clip,width=8.7cm]{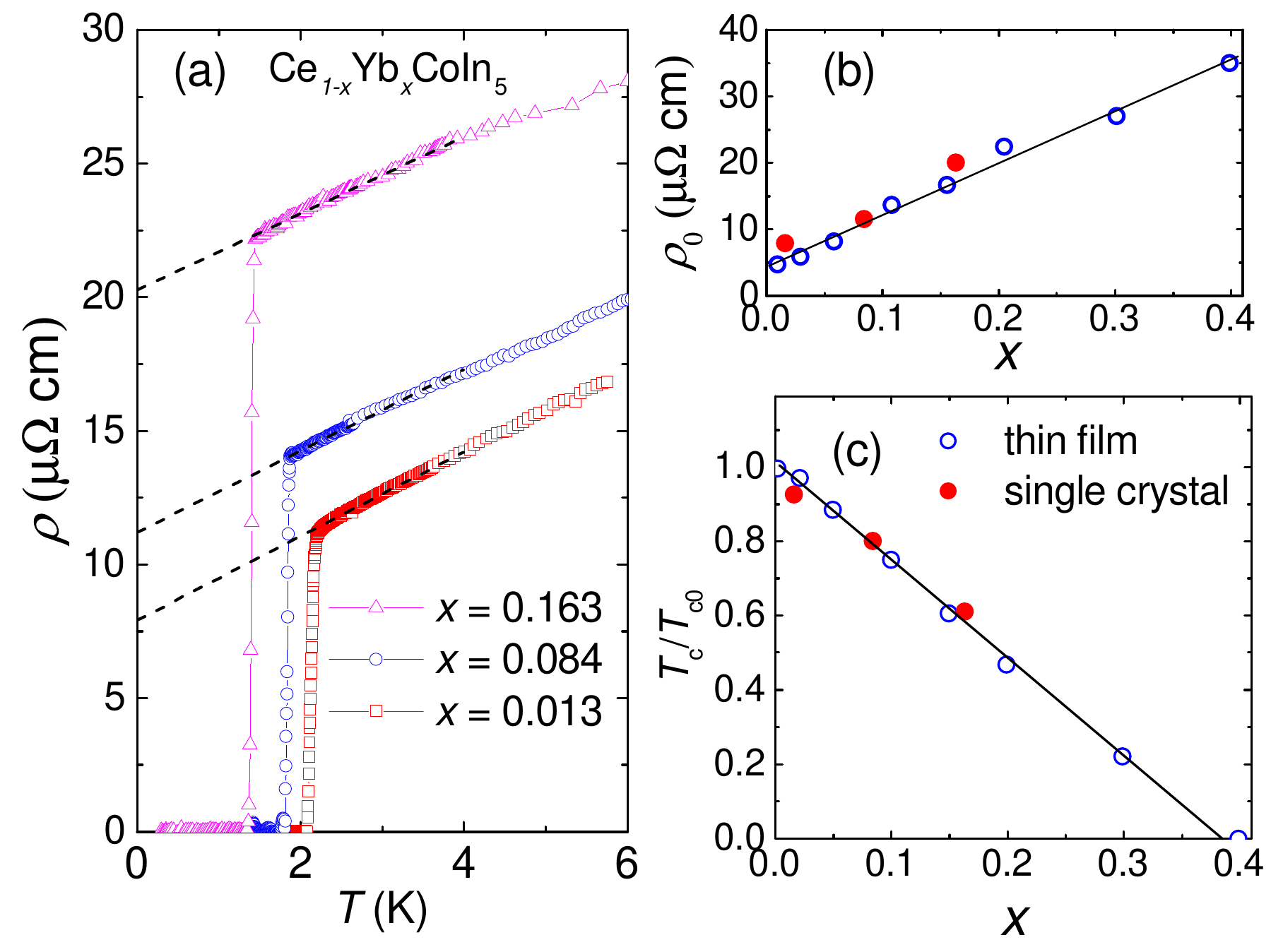}
\caption{(Color online). (a) Low-temperature resistivity of Ce$_{1-x}$Yb$_x$CoIn$_5$ single crystals. Here $x$ is the actual Yb concentration. The dashed lines are linear extrapolation to get the residual resistivity $\rho_0$. (b) and (c) Doping dependence of $\rho_0$ and normalized $T_c$/$T_{c0}$ ($T_{c0}$ = 2.3 K for pure CeCoIn$_5$) for our single crystals and the thin films (from Ref. \cite{Shimozawa}).}
\end{figure}


\begin{figure}
\includegraphics[clip,width=8.5cm]{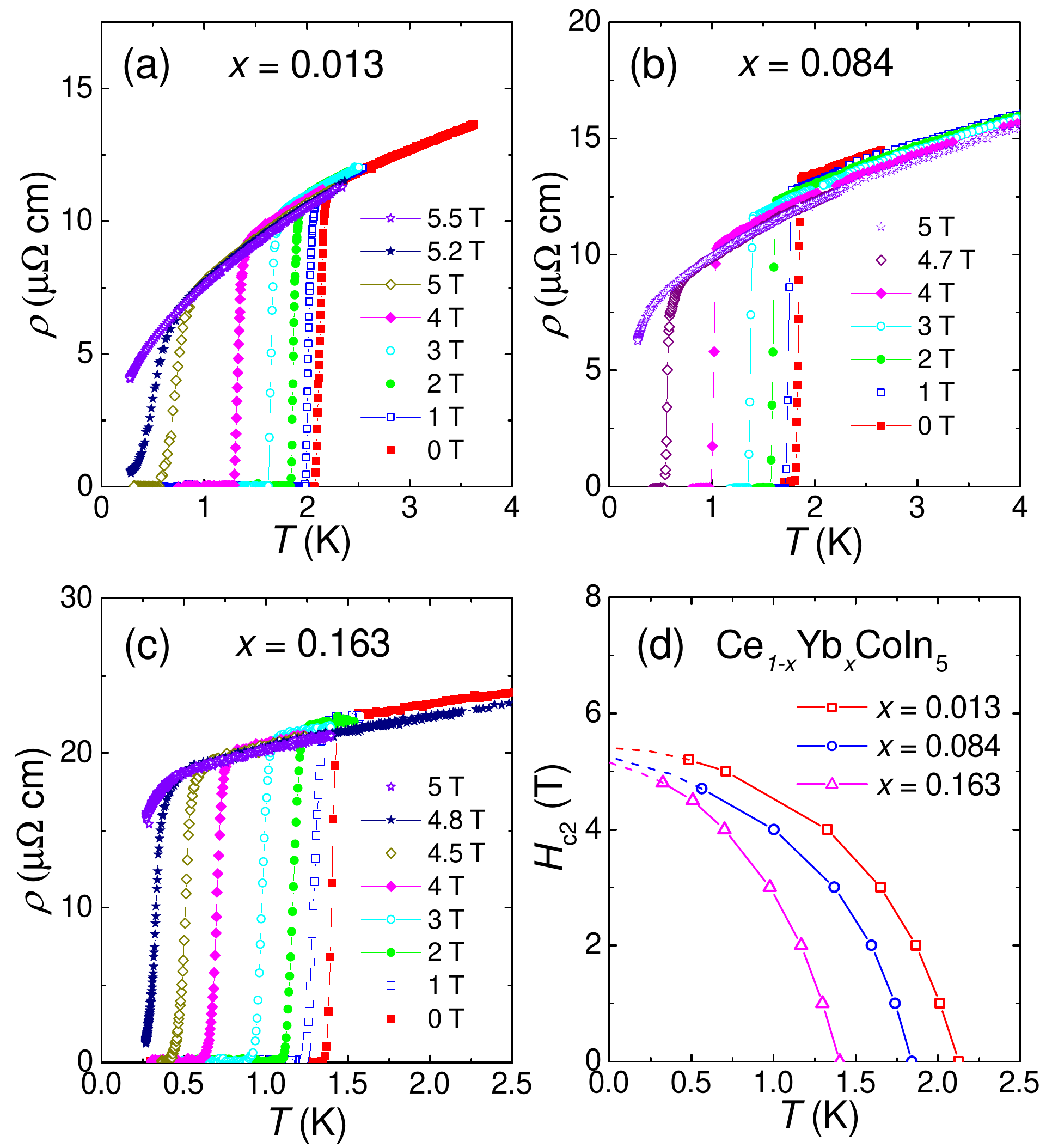}
\caption{(Color online). (a)-(c) Temperature dependence of resistivity for Ce$_{1-x}$Yb$_x$CoIn$_5$ single crystals at various magnetic fields $H \parallel c$ up to 5.5 T. (d) Temperature dependence of the upper critical field $H_{c2}(T)$ for all three samples.}
\end{figure}


Figure 1(a) shows the low-temperature in-plane resistivity $\rho(T)$ of Ce$_{1-x}$Yb$_x$CoIn$_5$. For $x$ = 0.013, 0.084, and 0.163, the normal-state resistivity $\rho(T)$ below 4 K is roughly linear, and their residual resistivity $\rho_0$ = 7.9, 11.2, and 20.2 $\mu\Omega$ cm are obtained by linear extrapolation. The $T_c$, defined as the midpoint of each resistive transition, is 2.13, 1.84, and 1.40 K, respectively. Figure 1(b) and 1(c) plot the doping dependence of $\rho_0$ and normalized $T_c/T_{c0}$ ($T_{c0}$ = 2.3 K for pure CeCoIn$_5$), respectively. For comparison, the data of Ce$_{1-x}$Yb$_x$CoIn$_5$ thin films from Ref. \cite{Shimozawa} are also plotted. The curves of single crystal and thin film nearly overlap with each other. This suggests that our determination of the actual Yb concentrations for our single crystals is accurate, and further confirms the conclusion of Ref. \cite{Jang}. According to Ref. \cite{Shimozawa}, the suppression of $T_c$ can be well reproduced by the AG pair breaking curve, suggesting that Yb ions act as impurity centers with unitary scattering, regardless of its valence.



Figure 2(a)-(c) present the resistivity $\rho(T)$ of Ce$_{1-x}$Yb$_x$CoIn$_5$ single crystals under various magnetic fields. Negative magnetoresistance and sub-$T$-linear $\rho(T)$ are observed in the normal state, as in pure CeCoIn$_5$ \cite{Paglione03}. To determine the zero-temperature upper critical field $H_{c2}(0)$, we plot the temperature dependence of $H_{c2}(T)$ in Fig. 2(d). With rough extrapolation, $H_{c2}(0) \approx$ 5.4, 5.2, and 5.1 T are obtained for $x$ = 0.013, 0.084, and 0.163, respectively. $H_{c2}(0)$ only exhibits slight decrease with the increase of Yb concentration, in contrast with the strong suppression of $T_c$.


The low-temperature in-plane thermal conductivity of Ce$_{1-x}$Yb$_x$CoIn$_5$ single crystals is shown in Fig. 3(a)-(c). In zero field, all the curves are roughly linear below 0.4 K, with a moderate slope. Previously for pure CeCoIn$_5$ at this temperature range, the slope is about 30 times larger than our doped samples, and constantly changing \cite{Movshovich,Louis,Flouquet}, which impedes an accurate extrapolation of $\kappa/T$ to zero temperature. $\kappa_0/T$ $<$ 2 mW K$^{-2}$ cm$^{-1}$ and $<$ 3 mW K$^{-2}$ cm$^{-1}$ was estimated for pure CeCoIn$_5$ in Refs. \cite{Movshovich} and \cite{Flouquet}, respectively. Here, due to the moderate slope of our doped samples, we linearly extrapolate $\kappa/T$ to zero temperature to obtain $\kappa_0/T =$ 1.22, 1.19, and 1.08 mW K$^{-2}$ cm$^{-1}$ for $x$ = 0.013, 0.084, and 0.163, respectively. These values are listed in Table I.


\begin{figure}
\includegraphics[clip,width=8.5cm]{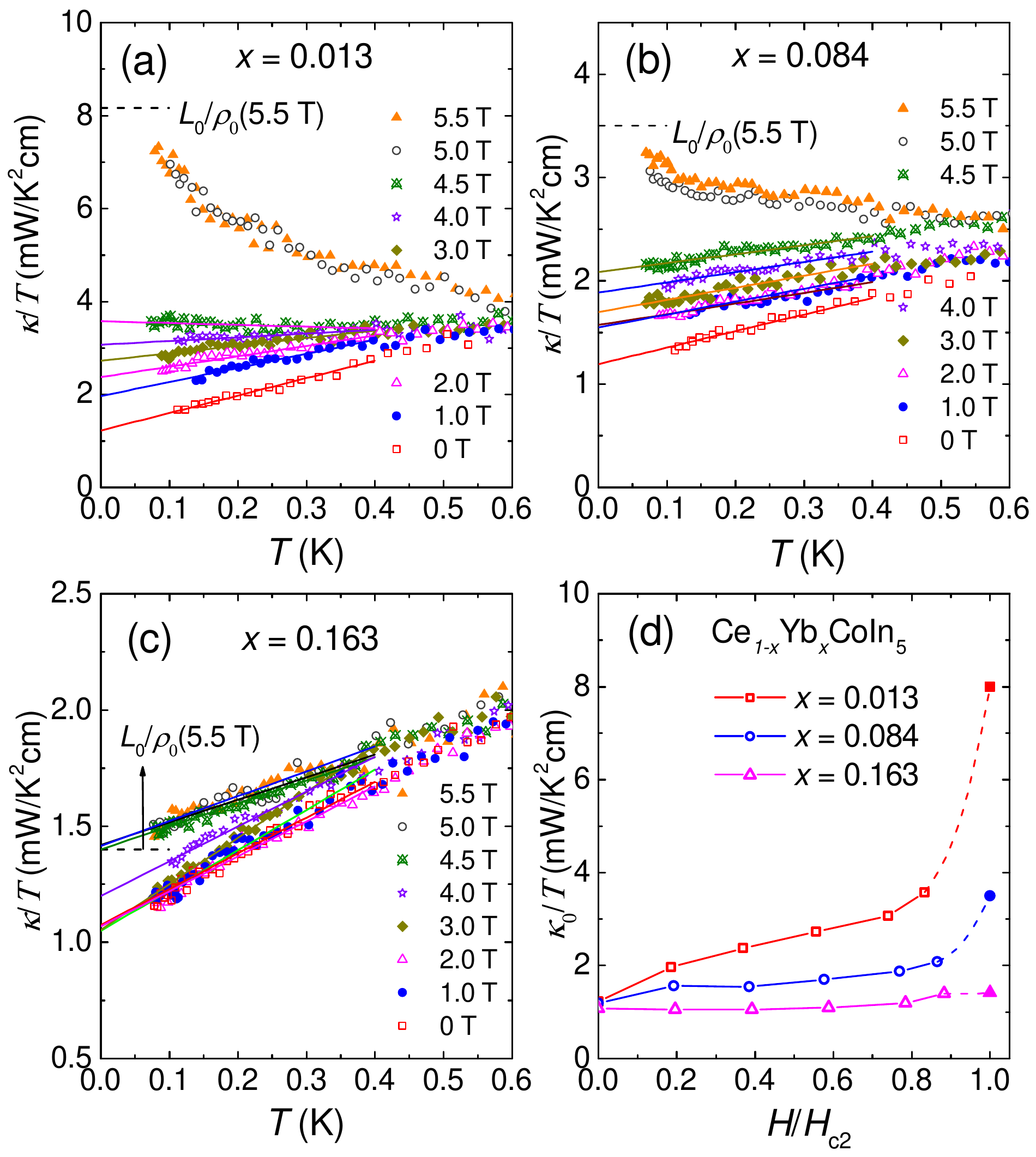}
\caption{(Color online). (a)-(c) Temperature dependence of the thermal conductivity divided by temperature $\kappa(T)/T$ of Ce$_{1-x}$Yb$_x$CoIn$_5$ under various magnetic fields. The solid lines are linear fits to extrapolate the residual linear term. The dashed lines indicate the normal-state Wiedemann-Franz law expectations $L_0$/$\rho_0$($H_{c2}$), with the Lorenz number $L_0$ = 2.45 $\times$ 10$^{-8}$ W$\Omega$K$^{-2}$. (d) Field dependence of $\kappa_0/T$. The field is normalized by the $H_{c2}$ of the three samples, respectively.}
\end{figure}


In magnetic fields, linear extrapolations still apply, except for the $x$ = 0.013 and 0.084 samples when $H >$ 4.5 T, as seen in Fig. 3(a)-(c). The $H =$ 5 and 5.5 T curves for $x$ = 0.013 and 0.084 samples tend to point to their normal-state Wiedemann-Franz law expectations $L_0$/$\rho_0$($H_{c2}$). The field dependence of $\kappa_0/T$ is shown in Fig. 3(d). For $x$ = 0.013, $\kappa/T$ increases gradually with field, followed by a jump to the normal-state value at $H_{c2}$. Such a jump of $\kappa_0/T$ near $H_{c2}$ was previously observed in pure CeCoIn$_5$, which was interpreted as the sign of a first-order superconducting transition \cite{Flouquet}. For $x$ = 0.084 and 0.163, this jump becomes less and less pronounced.

\begin{table}
\centering \caption{The properties of Ce$_{1-x}$Yb$_x$CoIn$_5$ and CeCo(In$_{1-y}$Cd$_y$)$_5$ single crystals. The actual Yb and Cd concentration $x_{act}$ and $y_{act}$ were determined from the WDS analysis. $T_{\rm c}$ is defined as the midpoint of the resistive transition.}\label{1}

\begin{center}
 \begin{tabularx}{0.45\textwidth}{p{1.3cm}p{1.3cm}p{1.3cm}p{1.6cm}p{1.6cm}}\hline\hline
$x_{nom}$ &~~~~ $x_{act}$ &~~~~~~~~~ $T_c $~(K)&~~~~~~~~~~~ $\kappa_{0}/T$~(mW/K$^2$cm) \\ \hline
0.05 &~~~ 0.013 &~~~~~~~~~~~ 2.13   &&1.22  \\
0.2 &~~~ 0.084 &~~~~~~~~~~~ 1.84    && 1.19     \\
0.4 &~~~ 0.163  &~~~~~~~~~~~ 1.40   && 1.08   \\ \hline
$y_{nom}$ &~~~~ $y_{act}$ &~~~~~~~~~ $T_c $~(K) &~~~~~~~~~~~ $\kappa_{0}/T$~(mW/K$^2$cm) \\ \hline
0.05 &~~~ 0.004  &~~~~~~~~~~~ 2.14   && 1.03   \\
0.075 &~~~ 0.008  &~~~~~~~~~~~ 2.05   && 0.90   \\
0.1 &~~~ 0.011  &~~~~~~~~~~~ 1.92   && 0.93   \\
\hline \hline
  \end{tabularx}
 \end{center}

\end{table}

\begin{figure}
\includegraphics[clip,width=5.8cm]{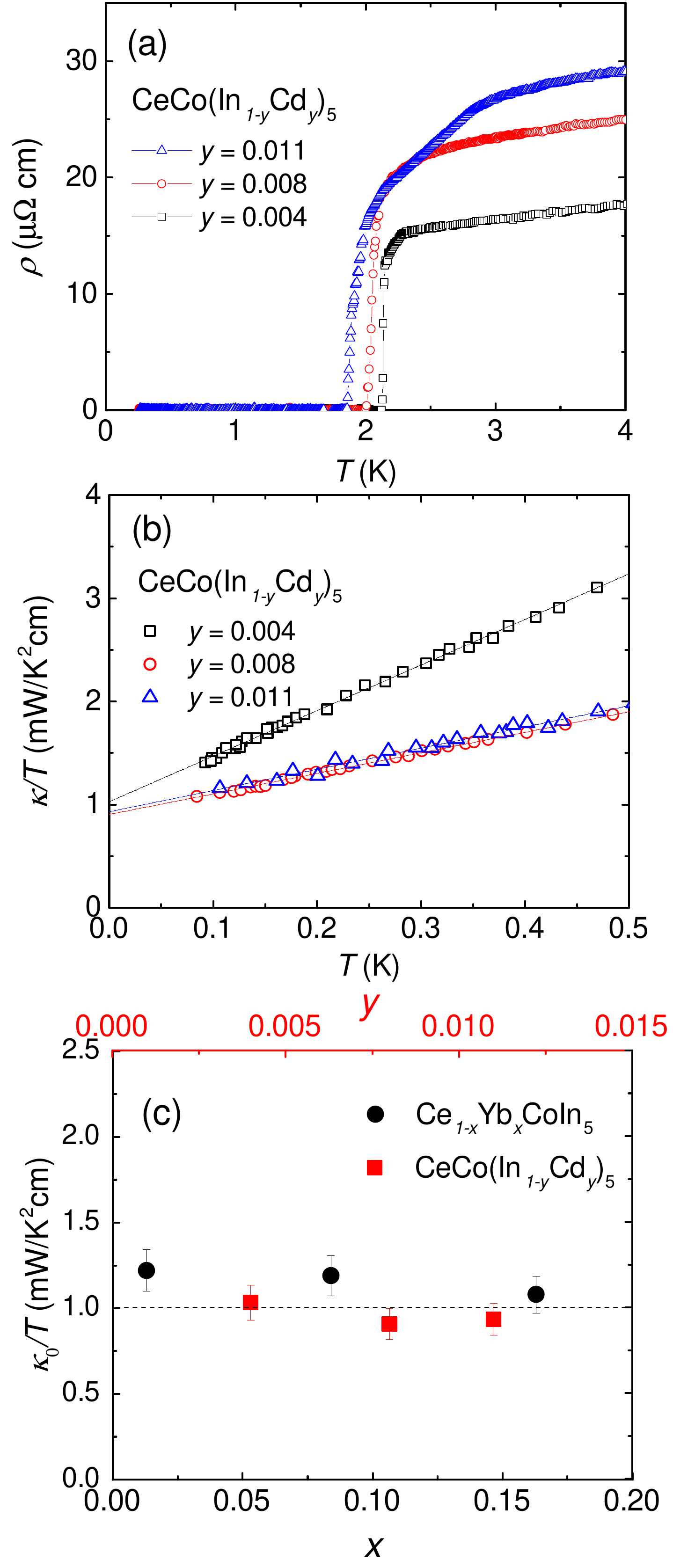}
\caption{(Color online). (a) Low-temperature resistivity of CeCo(In$_{1-y}$Cd$_y$)$_5$ single crystals. Here $y$ is the actual Cd concentration. (b) Temperature dependence of zero-field thermal conductivity divided by temperature $\kappa(T)/T$ for CeCo(In$_{1-y}$Cd$_y$)$_5$. The solid lines are linear fits to extrapolate $\kappa_0/T$. (c) $\kappa_0/T$ vs doping for all Ce$_{1-x}$Yb$_x$CoIn$_5$ and CeCo(In$_{1-y}$Cd$_y$)$_5$ samples at zero field. The error bar is determined from uncertainties on the geometric factor and the fit. The horizontal dashed line is the theoretical universal value for CeCoIn$_5$ with 2D nodal $d$-wave superconducting gap \cite{Movshovich}.}
\end{figure}

The major result of this work is that we observe a finite $\kappa_0/T$ with comparable values at zero field for all three Ce$_{1-x}$Yb$_x$CoIn$_5$ ($x$ = 0.013, 0.084, and 0.163) samples. To check whether this finite $\kappa_0/T$ also presents in doped CeCoIn$_5$ with other dopants, we measure the thermal conductivity of CeCo(In$_{1-y}$Cd$_y$)$_5$ ($y$ = 0.004, 0.008, and 0.011) single crystals. The low-temperature resistivity of CeCo(In$_{1-y}$Cd$_y$)$_5$ single crystals is shown in Fig. 4(a), from which $T_c$ = 2.14, 2.05, and 1.92 K are obtained, respectively. Figure 4(b) plots $\kappa/T$ vs $T$ at zero field for all three samples, where linear fit is used to get $\kappa_0/T$. The value of $\kappa_0/T$ is 1.03, 0.90, and 0.93 mW K$^{-2}$ cm$^{-1}$ for $y$ = 0.004, 0.008, and 0.011, respectively, as listed in Table I.

The $\kappa_0/T$ vs doping for all Ce$_{1-x}$Yb$_x$CoIn$_5$ and CeCo(In$_{1-y}$Cd$_y$)$_5$ samples at zero field is plotted in Fig. 4(c). Usually, the presence of a finite $\kappa_0/T$ is a strong evidence for nodal superconducting gap \cite{HShakeripour}. For example, $\kappa_0/T$ = 1.41 mW K$^{-2}$ cm$^{-1}$ for the overdoped $d$-wave cuprate superconductor Tl$_2$Ba$_2$CuO$_{6+\delta}$ (Tl-2201, $T_c$ = 15 K) \cite{Tl2201}, and $\kappa_0/T$ = 17 mW K$^{-2}$ cm$^{-1}$ for the $p$-wave superconductor Sr$_2$RuO$_4$ ($T_c$ = 1.5 K) \cite{SrRuO}. Although there is some uncertainty on the accurate value of $\kappa_0/T$ for pure CeCoIn$_5$ \cite{Movshovich,Flouquet}, the significant $\kappa_0/T$ observed here for all Yb and Cd doped CeCoIn$_5$ samples is quite reliable due to the moderate slope. Therefore, the nodal gap in Ce$_{1-x}$Yb$_x$CoIn$_5$ system persists at least up to $x$ = 0.163. This is at odds with the earlier penetration depth study \cite{Kim}. In Ref. \cite{Kim}, Yb doping leads to $n > 3$ ($\Delta\lambda(T) \sim T^n$) for $x_{nom}$ = 0.2 ($x \approx$ 0.04 determined by them), which suggests a nodeless superconducting gap \cite{Kim}. The reason for this discrepancy is not clear to us.

Furthermore, in Fig. 4(c), $\kappa_0/T$ manifests a nearly constant value around 1 mW K$^{-2}$ cm$^{-1}$ irrespective of Yb or Cd concentration, which demonstrates a universal heat conduction in Yb and Cd doped CeCoIn$_5$. The universal heat conduction is an important property of nodal $d$-wave superconducting gap, which means that the thermal conductivity is unaffected by change in the impurity scattering rate $\gamma$ \cite{Lee,Graf}. The universality results from the cancelation between two factors: (i) the density of Andreev bound states, which is proportional to $\gamma$, and (ii) the reduction of phase space for scattering of gapless excitations, which is proportional to $\gamma^{-1}$ \cite{Graf}. Experimentally, the universal $\kappa_0/T$ was observed in optimally doped high-$T_c$ cuprates YBa$_2$Cu$_3$O$_{6.9}$ and Bi$_2$Sr$_2$CaCu$_2$O$_8$ with $d$-wave gap \cite{Taillefer,Nakamae}. For pure CeCoIn$_5$ in which 2D nodal $d$-wave gap was also well-established, the universal $\kappa_0/T$ was theoretically estimated to be $\sim$1 mW K$^{-2}$ cm$^{-1}$ \cite{Movshovich}. This agrees very well with our experimental results of Ce$_{1-x}$Yb$_x$CoIn$_5$ and CeCo(In$_{1-y}$Cd$_y$)$_5$ systems, showing that the nodal $d$-wave superconducting gap in CeCoIn$_5$ is robust upon Yb and Cd doping. In this context, there is no need to propose a fully-gapped $d$-wave molecular superfluid of composite pairs beyond $x_{nom}$ = 0.2 \cite{Coleman}. Note that for the CeCo(In$_{1-y}$Cd$_y$)$_5$ system, despite that the substitution of Cd for In will introduce holes and the samples with $y$ = 0.008 and 0.011 have already entered the region where superconductivity and antiferromagnetism coexist \cite{Pham,Nicklas,Urbano}, the universal heat conduction still holds.

In summary, the heat transport properties of Ce$_{1-x}$Yb$_x$CoIn$_5$ and CeCo(In$_{1-y}$Cd$_y$)$_5$ systems have been systematically studied. We observe a finite value of $\kappa_0/T$ for $x$ up to 0.163 and $y$ up to 0.011. Furthermore, $\kappa_0/T$ is universal for both systems, with a value around 1 mW K$^{-2}$ cm$^{-1}$ which agrees very well with the theoretical estimation. These results demonstrate that the nodal $d$-wave superconducting gap in CeCoIn$_5$ is robust against Yb or Cd doping.

This work is supported by the Ministry of Science and Technology of China (National Basic Research Program No: 2012CB821402 and 2015CB921401), the Natural Science Foundation of China, China Postdoctoral Science Foundation No: 2014M560288, Program for Professor of Special Appointment (Eastern Scholar) at Shanghai Institutions of Higher Learning, and STCSM of China (No. 15XD1500200). Research at UCSD was supported by the U. S. Department of Energy, Office of Basic Energy Sciences, Division of Materials Science and Engineering under Grant No. DE-FG02-04-ER46105 (materials synthesis) and the National Science Foundation under Grant No. DMR 1206553 (materials characterization). Work at Brookhaven is supported by the US DOE under Contract No. DE-AC02-98CH10886. \\

${^\S}$ Present address: CNAM, Department of Physics, University of Maryland, College Park, Maryland 20742, USA.

${^\P}$ Present address: School of Materials Science and Engineering, Tianjin University of Technology, Tianjin 300384, China.

$^*$ jkdong@fudan.edu.cn

${^\dag}$ leishu@fudan.edu.cn

${^\ddag}$ shiyan$\_$li@fudan.edu.cn

\end{document}